\documentclass[aps,pre,one,11pt]{revtex4}%
\usepackage{epsfig}
\usepackage{amsmath}
\usepackage{graphicx}

\newcommand{\degree}{$^\circ$}
\newcommand{\pt}{\perp}
\newcommand{\pa}{{\scriptscriptstyle \|}}

\begin{document}

\title{\LARGE{Maxwell and Kerr effects in solutions of macromolecules with dendrons in side groups}}

\author{Nikolai~Tsvetkov${^1}$}

\email{ntsvet@mail.wplus.ru}

\author{Larisa~Andreeva${^2}$}
\author{Irina~Strelina${^2}$}
\author{Tatyana~Dmitrieva${^1}$}
\author{Ilya~Martchenko${^1}$}
\author{Nina~Girbasova${^3}$}
\author{Alexander~Bilibin${^3}$}

\affiliation{${^1}$ Department of Physics, St Petersburg State University, Ul. Ulyanovskaya 3, 198504 Stary Peterhof, Russia}
\affiliation{${^2}$ Institute of Macromolecular Compounds, Russian Academy of Sciences, Bolshoi pr. Vas. Ostr. 31, 199004 St Petersburg, Russia}
\affiliation{${^3}$ Department of Chemistry, St Petersburg State University, Universitetsky pr. 26, 198504 Stary Peterhof, Russia}

\begin{abstract}

Linear polymers with dendrons of first and second generations based on $L$-aspargic acid are studied by the methods of flow birefringence and electric birefringence. Optical, dipole, dynamic, and conformational properties of the macromolecules, as well as contributions to their optical anisotropy due to macroform and microform effects, are investigated. The polymers are found to possess permanent electric dipole moments and undergo reorientation in external electric and hydrodynamic fields according to large-scale rotation mechanism. A significant growth in equilibrium rigidity, optical anisotropy, and dipole moment of monomer units is observed when rigid benzamide fragments are introduced into the dendrimers. The backbone conformation of second-generation dendrimers containing such fragments is shown to be well described by generalized wormlike chain model.

\vspace{4 mm}

The paper was presented at the \textit{4th All-Russian Kargin Conference} (Moscow State University, January 29 -- February 2, 2007) with abstracts published in conference proceedings, Vol.~{\bf 3}, 65. A revised version appeared in Polymer Sci. A, {\bf 50}, No. 2, 119 (2008).

\end{abstract}

\maketitle

\section{\label{sec:intr} Introduction}

The interest in dendritic macromolecules is associated with intense research of hyperbranched polymer structures. Dendrimers attract a special attention due to opportunities of modifying their surface with various functional groups and finding new promising practical applications.

\begin{figure} [!ht]
\centering
\includegraphics[scale=0.7]{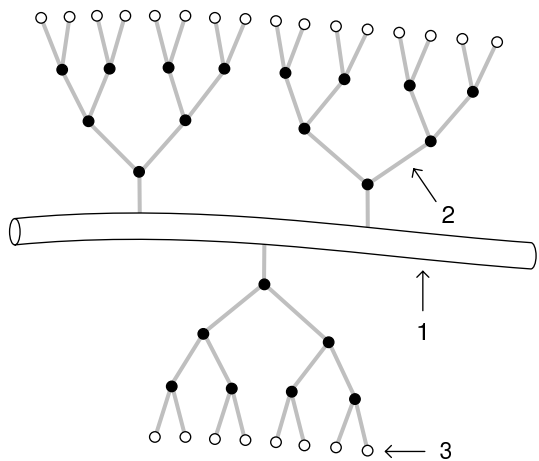}
\caption{Cylindrical dendrimer: main polymer chain (1), side dendrons (2), terminal groups (3)}
\label{cyl}
\end{figure}

Earlier research \cite{1,2,3} has shown that the properties of linear dendronized polymers (Fig.~\ref{cyl}) are strongly dependent on the generation number of side dendritic substituents. However, the role of structural parameters of dendrons in the formation of properties of macromolecules has not yet been sufficiently tested.

Systematic investigations of linear polyacrylates with side dendrons based on $L$-aspargic acid have been performed in \cite{4,5,6,7,8,9}. A lack in optically anisotropic groups was responsible for quite low optical anisotropy of these polymers. A significant number of hydrogen bonds between side substituents provided an unusual combination of low equilibrium rigidity and considerable kinetic rigidity.

Two series of polymers under study (P1 and P2) had differences in chemical structure, with P2 bearing rigid benzamide fragments in side substituents. Both series included first-generation (P1-1, P2-1) and second-generation (P1-2, P2-2) samples of various polymerization degrees.

\begin{figure}[!ht]
\centering
\includegraphics{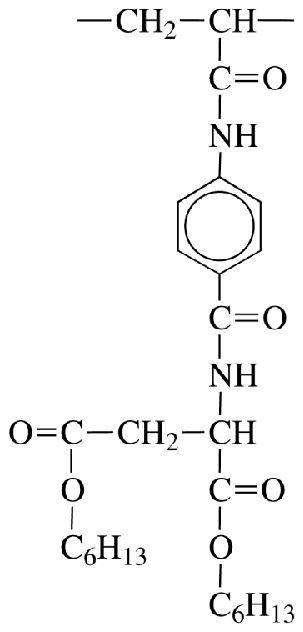}
\caption{P3-1}
\label{05}
\end{figure}

\begin{figure}[!ht]
\centering
\includegraphics{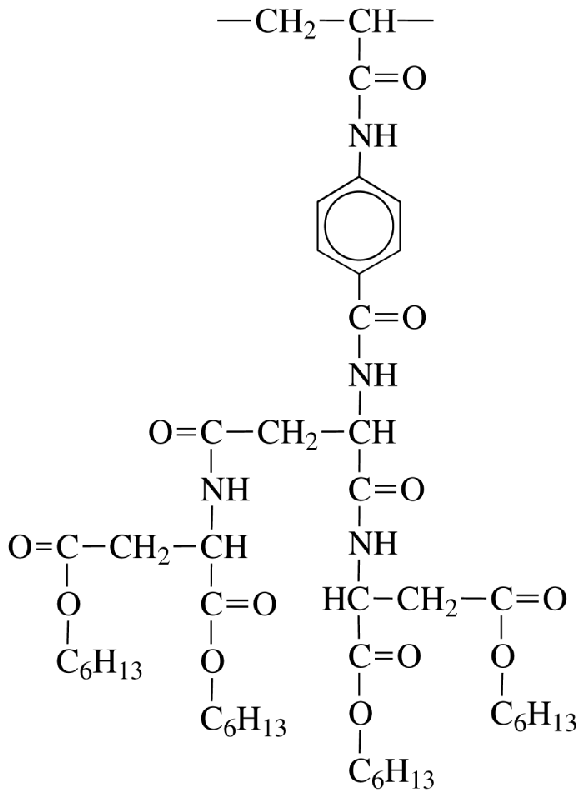}
\caption{P3-2}
\label{06}
\end{figure}

The recently synthesized P3 series of dendrimers contain long aliphatic terminal groups $\mathrm C_6\mathrm H_{13}$ in side dendritic substituents (Fig.~\ref{05}, Fig.~\ref{06}). In \cite{9}, the hydrodynamic properties of these macromolecules have been investigated.

Besides making macromolecules more branched, the introduction of hexylocarbonyl groups significanty improves the solubility of the polymers. This effect makes it possible to study the first-generation and second-generation dendrimers of the P3 series (P3-1, P3-2) not only in the mixture of dichloroacetic acid (DCA) with 0.3~M LiCl, but also in bromoform.

In this study, the dendrimers of the P3 series are investigated by the method of dynamic birefringence (the Maxwell effect) and the method of equilibrium and non-equilibrium electric birefringence (the Kerr effect.)

\section{\label{sec:expt} The Experiment}

The dynamic flow birefringence (the Maxwell effect) was investigated in a dynamooptimeter equipped with 3-$cm$-diameter and 3.21-$cm$-height internal rotor. The gap between the stator and rotor was $0.022~cm$. To enhance sensitivity, the photoelectric registration scheme with modulation of ellipticity of light polarization \cite{10} was used; a He-Ne laser operating at wavelength $\lambda = 632.8~nm$ was employed as a light source. The elliptic rotary compensator had a relative optical path difference $\Delta\lambda/\lambda = 0.0366$. The dynamic flow birefringence was examined at 24\degree \ with forced water thermostating.

Equilibrium and non-equilibrium birefringence was studied in rectangular-pulsed \cite{11} and sinusoidal-pulsed \cite{12} electric fields. To increase sensitivity, the photoelectric registration scheme with modulation of light polarization ellipticity was employed \cite{10,13}. The He-Ne laser with operation wavelength $\lambda = 632.8~nm$ was used as a light source. The relative optical path difference of the rotating elliptic compensator was $\Delta\lambda/\lambda = 0.01$. The measurements were performed in a glass Kerr cell with titanium electrodes $3~cm$ in length along the light beam; the gap between electrodes was $0.03~cm$.

\section{\label{sec:res} Results and Discussion}

\subsection{Flow Birefringence}

The anisotropy of a chain molecule \cite{10} is determined by the optical polarizability of a segment ($\alpha_1-\alpha_2$) and the degree of coiling of the molecule $h/L$:
\begin{equation}
(\gamma_1 - \gamma_2)_i = \frac{3}{5}(\alpha_1 - \alpha_2)\frac{h^2}{AL}/ \left[1-\frac{2h^2}{5 L^2}\right],
\label{eq01}
\end{equation}
where ($\alpha_1-\alpha_2$) is difference in the polarizabilities of a segment, $h$ is distance between chain ends, $A$ is statistical segment length, and $L$ is contour length of the chain. The segmental anisotropy of a macromolecule ($\alpha_1 - \alpha_2$) is defined by the difference in polarizabilities of the monomer unit ($\alpha_{\pt} - \alpha_{\pa}$) in perpendicular and parallel chain directions and the number of monomer units $S$ in the Kuhn segment $(\alpha_1 - \alpha_2) = (\alpha_{\pt} - \alpha_{\pa})S$, $S = A/\lambda$ (where $\lambda$ is length of the monomer unit).

If the refractive index of a solvent $n_s$ differs from that of a dissolved polymer $n_k$, additional anisotropies due to optical interactions between separate portions of a macromolecule should be taken into account.

The anisotropy of the macroform of a chain macromolecule is defined by
\begin{equation}
(\gamma_1 - \gamma_2)_f = \frac{\left(n_k^2 - n_s^2\right)^2}{\left(4 \pi n_s \rho N_A \right)^2}\frac{M^2}{V}\left(L_2 - L_1\right),
\label{eq02}
\end{equation}
where $\rho$ is density of the dry polymer, $M$ is its molecular mass, $V$ is volume occupied by a macromolecule in solution, and ($L_2 - L_1$) is parameter depending on the shape asymmetry of the macromolecule.

The anisotropy of the microform of a macromolecule may be described by relationship
\begin{equation}
(\gamma_1 - \gamma_2)_{f_s} = \frac{3(n_k^2 - n_s^2)^2}{5(4 \pi n_s)^2}\frac{M_A}{\rho N_A} \frac{h^2}{AL}(L_2 - L_1)_s
\label{eq03}
\end{equation}

Here, $M_A$ is molecular mass of the segment and $(L_2-L_1)_s$ is parameter depending on the segment asymmetry.

The dynamic birefringence of the P3-1 and P3-2 series was studied in solutions in bromoform and dichloroacetic acid containing small amounts of LiCl. The measurements were carried out in solvents with different refractive indexes $n_s$ and $n_k$. The refractive indexes $n_s$ and the
refractive index increments $dn/dc$ (average values for the samples) for P3-1 and P3-2 are shown in Table 1. The samples under study are characterized by high molecular masses and high masses of statistical segments \cite{9}.

\begin{table}
$$\begin{array}{|c|l|c|c|}
\hline
\text{Polymer} & \multicolumn{1}{c|}{\text{Solvent}} & n_s & dn/ds \\
\hline
\text{P3-1} & Bromoform & 1.586 & -0.054 \\
            & DCA + 0.3 M LiCl & 1.468 & 0.064 \\
\text{P3-2} & Bromoform & 1.586 & -0.079 \\
            & DCA + 0.3 M LiCl & 1.468 & 0.039 \\
\hline
\end{array}$$
\caption{Refractive indexes and increments for the systems under study}
\end{table}

\begin{figure}[!ht]
\centering
\includegraphics{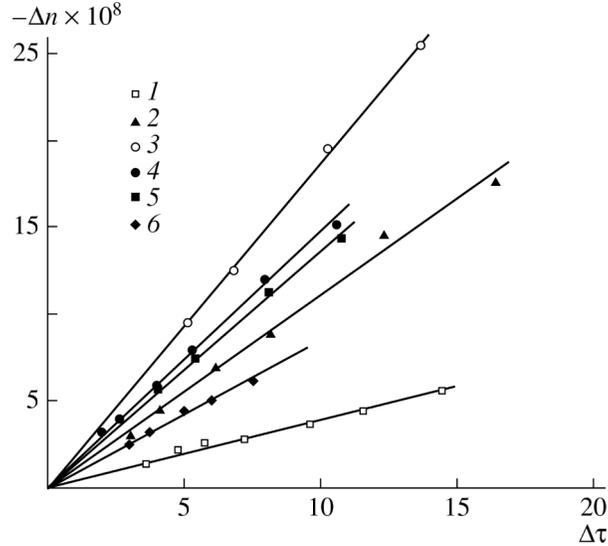}
\caption{Dependence of the birefringence $\Delta n$ on the shear stress $\Delta \tau$ for P3-2-1 in bromoform~(1), and P3-1-6~(2), P3-2-4~(3 -- 5), and P3-2-6~(6) in DMA + 0.3 M LiCl at concentrations $c \times 10^2 = $ 0.94~(1), 0.32~(2), 1.0~(3), 0.75~(4), 0.38~(5), and 0.25~(6) $g/cm^3$.}
\label{07}
\end{figure}

Therefore, contributions due to macroform and microform effects to the optical anisotropy of the chain molecule cannot be neglected. Fig.~\ref{07} illustrates the experimental relationships between the flow birefringence $\Delta n$ and the shear stress $\Delta \tau = g(\eta - \eta_0)$, where $g$ is shear rate, and $\eta$ and $\eta_0$ are the viscosities of solution and solvent, respectively, for a number of P3 solutions.

In the region of small shear stresses, these relationships are linear and pass through the origin. That provides evidence of the molecular dispersity of the solutions under study. The optical shear coefficients $\Delta n / \Delta \tau $ calculated from the slopes of straight lines in Fig.~\ref{08} are negative and large with respect to their absolute values. Table 2 lists the optical shear optical coefficients
\begin{equation}
\frac{[n]}{[\eta]} = \lim_{\substack{g\to 0 \\ c\to 0}} \frac{\Delta n}{\Delta \tau}
\end{equation}
measured for P3-1 and P3-2 samples in bromoform and in the DCA + 0.3~M LiCl mixture.

\begin{table}
$$\begin{array}{c|c|c|@{}c@{}|@{}c@{}|c|c|c}
\hline
\text{Sample} & M_{D\eta}\times 10^{-3} & N & \multicolumn{2}{c|}{[n]/[\eta] \times 10^{10}, s^2cm/g} & \tau \times 10^6, s & F & K{\times}10^{10}, g^{-1} cm^5 (300 V)^{-2} \\
\cline{4-5} \cline{6-8}
& & & Bromoform & DCA + 0.3 M LiCl & \multicolumn{3}{c}{Bromoform} \\
\hline
\multicolumn{8}{c}{\text{Polymer P3-1}} \\
1* & 3893* & 8445* & -36*   &      & 80* & 1.38* & -50*  \\
2  & 1039 & 2384 & -46.6 & -100 & 63 & 0.39 & -145 \\
3  & 937  & 2033 & -40   &      & 76 & 0.24 & -146 \\
4  & 597& 1295 & -46.2  & 100  & 48 & 0.19 & -91   \\
5  & 585& 1270 & -40   &      & 25 & 0.35 & -100 \\
6  & 390  & 846  &      & -111 & 36 & 0.12 & -87  \\
7  & 130& 281  & -21   &      & 6.3& 0.10 & -23  \\
\multicolumn{8}{c}{\text{Polymer P3-2}} \\
1  & 1056 & 1229 & -35.2 & -120 & 32 & 0.21 & -57  \\
2  & 764& 889  & -16.6 &      &22.5& 0.18 & -53   \\
3  & 733& 853  & -34   &      & 28 & 0.13 & -53  \\
4  & 545& 634  & -26   & -117  &    &      &      \\
5  & 511& 595  & -43.3 &      & 16 & 0.16 & -51  \\
6  & 89.4 & 104  &       & -86  & 9.2& 0.03 & -28  \\
7  & 30.5 & 36   &       &      &    &      & -3.6 \\
\hline
\end{array}$$
\caption{Molecular dynamics and electrooptical characteristics of P3-1 and P3-2 in bromoform and DCA + 0.3 M LiCl. An asterisk (*) indicates that dioxane was used as the solvent for this sample.}
\end{table}

For the systems with different values of $n_k$ and $n_s$, the optical shear coefficient $[n]/[\eta]$ may be written as a sum of three terms \cite{10}:

\begin{equation}
\frac{[n]}{[\eta]} = \frac{[n]_i}{[\eta]} + \frac{[n]_{f_s}}{[\eta]} + \frac{[n]_f}{[\eta]}.
\label{eq04}
\end{equation}

The first term is the contribution related to the inherent optical anisotropy of molecules $(\gamma_1 – \gamma_2)_i$. The second and third terms relate to contributions due to anisotropies of microform and macroforms, respectively.

If the chain is sufficiently coiled $(h \ll L)$ and undeformed, then factor $h^2/AL = \langle h^2 \rangle /AL$ in (\ref{eq01}) and (\ref{eq03}) is equal to unity. The first and second terms in formula (\ref{eq04}) may be expressed as \cite{10}:
\begin{equation}
\frac{[n]_i}{[\eta]} = \frac{4 \pi (n_s^2 + 2)^2}{45 k T n_s}(\alpha_1 - \alpha_2)
\label{eq05}
\end{equation}
(where $T$ is temperature, and $k$ is Boltzmann constant) and
\begin{equation}
\frac{[n]_{f_s}}{[\eta]} = \frac{(n_s^2 + 2)^2 (n_k^2 - n_s^2)}{180 \pi RT n_s^3 \rho} M_A (L_2 - L_1)_s.
\label{eq06}
\end{equation}

The values of $M_A$ were taken from \cite{9}. To determine the segment asymmetry parameter $(L_2-L_1)_s$, the expression (5.24) from \cite{10} and the results of conformational investigations from \cite{9} (Table 3) were used. An analysis of the hydrodynamic data within the framework of wormlike and generalized \cite{14} models shows that the axial ratio of segment $A/d$ for P3-2 is considerably smaller than that for P3-1 both in bromoform and the acidic solvent.

\begin{table}
$$\begin{array}{c|l|c|c|c|c|c|c|c|c}
\hline
\text{Polymer} & \multicolumn{1}{c|}{\text{Solvent}} & A,~\AA & d,~\AA & A/d & A,~\AA &
d,~\AA & A/d & \multicolumn{2}{c}{M_A \times 10^{-3}} \\ \cline{3-5}\cline{6-8}\cline{9-10}
& & \multicolumn{3}{c|}{WLM} & \multicolumn{3}{c|}{GM} & WLM & GM \\
\hline
\text{P3-1} & Bromoform       & 100 & 23 & 4.35 &     &    &     & 18.96 &       \\
     & DCA + 0.3 M LiCl  & 180 & 36 & 5    &     &    &     & 34.1  &       \\
\text{P3-2} & Bromoform       & 70  & 50 & 1.4  & 104 & 74 & 1.4 & 24.4  & 58.0  \\
     & DCA + 0.3 M LiCl  & 140 & 70 & 2    & 250 & 75 & 3.3 & 48.8  & 139.5 \\
\hline
\end{array}$$
\caption{Asymmetry and mass of segments for P3-1 and P3-2 \cite{9}. WLM is an abbreviation for the wormlike model, and GM is an abbreviation for the generalized model.}
\end{table}

The third term in (\ref{eq04}) is the macroform effect expressed as

\begin{equation}
\frac{[n]_{f}}{[\eta]} = \frac{0.058 \Phi (n_s^2 + 2)^2 (n_k^2 - n_s^2)}{\pi \rho^2 N_A RT n_s^3 \rho} \frac{M}{[\eta]}.
\label{eq07}
\end{equation}

The specific property of the macroform effect is its dependence on the molecular mass of the polymer. The value of the Flory coefficient $\Phi$ in (\ref{eq07}) equal to $1.5 \times 10^{23}$ was found from the molecular characteristics of highest molecular mass fractions \cite{9}.

As noted above, relationships (\ref{eq05}) -- (\ref{eq07}) are applicable for the coiled chains which obey Gaussian statistics.

Table 4 shows the contributions of macroforms and microforms for the highest molecular mass fractions of the samples. These contributions were calculated through relationships (\ref{eq06}) and (\ref{eq07}). For P3-2, whose hydrodynamic data were discussed in \cite{9} in terms of wormlike and generalized models, the contribution $[n]_{f_s}/[\eta]$ was estimated with the use of these two models. As follows from Table 4, contributions $[n]_f / [\eta]$ and $[n]_{f_s}/[\eta]$ are sufficiently large (this is especially true of $[n]_{f_s}/[\eta]$) and are comparable to the total effect.

\begin{table}

$$\begin{array}{@{}c@{}|@{}l@{}|c|@{}c@{}|@{}c@{}|@{}c@{}|@{}c@{}|@{}c@{}|@{}c@{}}
\hline
\text{Polymer} & \multicolumn{1}{c|}{\text{Solvent}} & [\eta], & [n]/[\eta] \times 10^{10}, & [n]_f / [\eta] \times 10^{10}, & \multicolumn{2}{@{}c@{}|}{[n]_{fs} / [\eta] \times 10^{10}, s^2cm/g  } & \multicolumn{2}{@{}c@{}}{[n]/[\eta]_i \times 10^{10}, s^2cm/g} \\ \cline{6-7}\cline{8-9}
 &  & dl/g & s^2cm/g (exp.) & s^2cm/g & WLM & GM & WLM & GM \\ [2pt]
\hline
\text{P3-1-2} & Bromoform      & 0.59 & -46.6 & 7.35 & 18.0 &       & -72.0 &        \\
\text{P3-1-3} & Bromoform      & 0.51 & -40   & 7.3  & 18.0 &       & -65.3 &        \\
\text{P3-1-4} & Bromoform      & 0.41 & -46.2 & 5.7  & 18.0 &       & -70.0    &        \\
\text{P3-1-2} & DCA + 0.3 M LiCl & 1.3  & -100  & 4.3  & 32.8 &       & -137.1&        \\
\text{P3-1-4} & DCA + 0.3 M LiCl & 0.60 & -100  & 3.6  & 32.8 &     & -136.4&        \\
\text{P3-1-6} & DCA + 0.3 M LiCl & 0.78 & -111  & 1.8  & 32.8 &     & -145.6&        \\
\text{P3-2-1} & Bromoform      & 0.17 & -35.2 & 38.0 & 11.8 & 28.5  & -85.0    & -101.7  \\
\text{P3-2-2} & Bromoform      & 0.14 & -16.6 & 33.0 & 11.8 & 28.5  & -61.4 & -78.1  \\
\text{P3-2-3} & Bromoform      & 0.13 & -34   & 32.9 & 11.8 & 28.5  & -78.7 & -95.4  \\
\text{P3-2-4} & Bromoform      & 0.14 & -26   & 23.5 & 11.8 & 28.5  & -61.3 & -78.0    \\
\text{P3-2-5} & Bromoform      & 0.13 &  -43.3 & 23.8 & 11.8 & 28.5  & -78.9 & -95.6   \\
\text{P3-2-1} & DCA + 0.3 M LiCl & 0.49 & -120  & 4.2  & 10.0 & 42.5  & -134.2& -167   \\
\text{P3-2-4} & DCA + 0.3 M LiCl & 0.43 & -117  & 1.8  & 10.0 & 42.5  & -128.8& -161.3 \\
\hline
\end{array}$$
\caption{Contributions of macroform and microform effects to the dynamic birefringence. WLM is an abbreviation for the wormlike model, and GM is an abbreviation for the generalized model.}
\end{table}

Above all, this phenomenon may be explained by an appreciable weight of the statistical segment $M_A$.

Table 5 lists the optical properties of P3-1 and P3-2 polymers (average values for the highest molecular mass samples) obtained with due regard for optical long-range and short-range actions along with the  data reported previously for P1-1, P1-2, P2-1, and P2-2. As follows from this table, the introduction of rigid benzamide fragments into P2 and P3 between the main chain and side dendrons leads to a drastic (by dozens of times) change in the optical characteristics of macromolecules relative to those of polymers of series P1. In this case, the equilibrium rigidity increases, but this effect is not so pronounced.

\begin{table}
$$\begin{array}{c|l|c|c|c|c}
\hline
\text{Polymer} & \multicolumn{1}{c|}{\text{Solvent}} & [n]/[\eta] \times 10^{10}, s^2cm/g & (\alpha_1 - \alpha_2) \times 10^{25}, cm^3 & S & (\alpha_1 - \alpha_2) \times 10^{25}, cm^3 \\
\hline
\text{P3-1} & Bromoform         & -64  & -790   & 40   & -19.8 \\
\text{P3-1} & DCA + 0.3 M LiCl  & -140 & -1730  & 72   & -24.0 \\
\text{P3-2} & Bromoform         & -73  & -860   & 28   & -30.7 \\
\text{P3-2} & Bromoform         & -90* & -1060* & 67*  & -15.8*\\
\text{P3-2} & DCA+0.3 M LiCl  & -132 & -1620  & 56   & -28.9 \\
\text{P3-2} & DCA+0.3 M LiCl  & -164*& -2025* & 160* & -12.6*\\
\text{P2-1} & \text{o-}Toluidine       & -28  & -320   &      &       \\
\text{P2-2} & \text{o-}Toluidine       & -60  & -690   &      &       \\
\text{P1-1} & Dioxane           & -2.0 & -25    & 12.5 & -2.0   \\
\text{P1-2} & Dioxane           & -2.5 & -30    & 18   & -2.5  \\
\hline
\end{array}$$
\caption{Optical characteristics of P3-1, P3-2, P2-1, and P2-2 in various solvents. An asterisk (*) indicates that the calculations rely on the interpretation of hydrodynamic data \cite{9} using the generalized model \cite{14}.}
\end{table}

A comparison between the results obtained for P2 and P3 polymers demonstrates that the lengthening of terminal groups in P3 leads to an increase in $[n]_{i}/[\eta]$ and, consequently, in the anisotropy of the statistical segment. To gain insight into the causes of this effect (a growth of the equilibrium rigidity or an increase in anisotropy of the
monomer unit), parameters of the equilibrium rigidity should be quantitatively estimated via an independent procedure. An attempt to obtain such data for the polymers of series P2 was not successful \cite{6}.

Transition from the dendrons of first generation (P3-1) to the second generation (P3-2) leads to the optical anisotropy of the monomer unit $(\alpha_{\pt}-\alpha_{\pa}) = (\alpha_1-\alpha_2)/S$ becoming more negative in both bromoform and dichloroacetic acid, when estimated in terms of the wormlike model.

The application of the generalized model for estimation of the equilibrium rigidity (parameter $S$) for P3-2 from the hydrodynamic data yields a twofold decrease in the absolute values of $(\alpha_{\pt}-\alpha_{\pa})$ in the case of both solvents as compared with the corresponding values derived in terms of the wormlike model. The contribution of the side dendron to the optical anisotropy of the chain is determined by the angle between the optical axis of dendron and the direction of the backbone growth. The generalized model assumes the imperfection of the planar structure of the backbone that may lead to spiralization of the maximally elongated conformation of the chain and, consequently, to a different orientation angle between the dendron optical axis and the backbone direction, if compared with the \textit{trans}-planar chain.

\subsection{Non-Equilibrium Electric Birefringence}

\begin{figure}[!ht]
\centering
\includegraphics{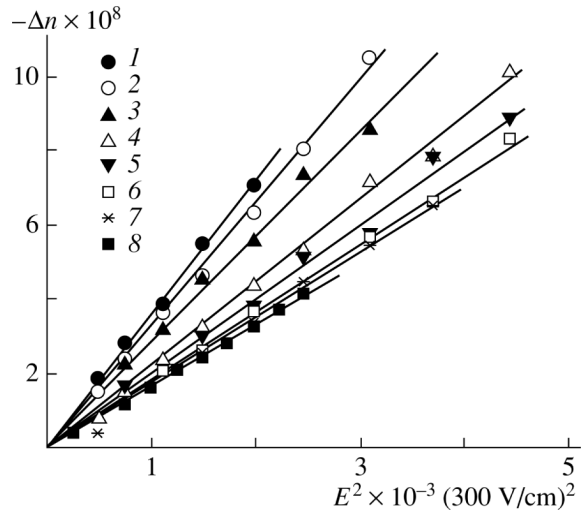}
\caption{Dependences of birefringence $\Delta n$ on the square of sinusoidal-pulsed electric field $E^2$ for the bromoform solution of polymer P3-1-7 at the concentration $c = 1.29 \times 10^{-2} g/cm^3$. The plot includes data for the frequencies $\nu =$ 8~(1), 13~(2), 21~(3), 50~(4), 100~(5), 200~(6), and 600~(7) kHz, and for the solvent~(8) at the same frequencies.}
\label{08}
\end{figure}

Non-equilibrium electric birefringence for P3-1 and P3-2 polymers has been investigated in bromoform, but a single sample, P3-1-1, has been studied in a non-polar solvent (dioxane). Fig.~\ref{08} illustrates the data obtained with the non-equilibrium Kerr effect method. Here, electric birefringence $\Delta n$ is plotted as a function of intensity of the sinusoidal-pulsed field $E^2$ for polymer P3-1-7.

The observed dependences are straight lines. This fact indicates that the Kerr law is fulfilled. The slope of experimental plots decreases with increasing frequency of the sinusoidal-pulsed field and approaches that of the solvent. This fact suggests the dispersion of electric birefringence in polymer solutions in the radio-frequency range. Similar dependences were observed for all polymers under study. That allows determination of the specific Kerr constant $K_{\nu}$ of a polymer at the specified frequency of the sinusoidal field $\nu$ via the expression

\begin{equation}
K_{\nu} = \frac{\Delta n_{\nu}}{c E^2}.
\end{equation}

Here, $\Delta n_{\nu}$ is electric birefringence contribution of the polymer at a frequency $\nu$, c is concentration of the studied solution, and $E$ is intensity of the electric field.

\begin{figure}[!ht]
\centering
\includegraphics[scale=0.9]{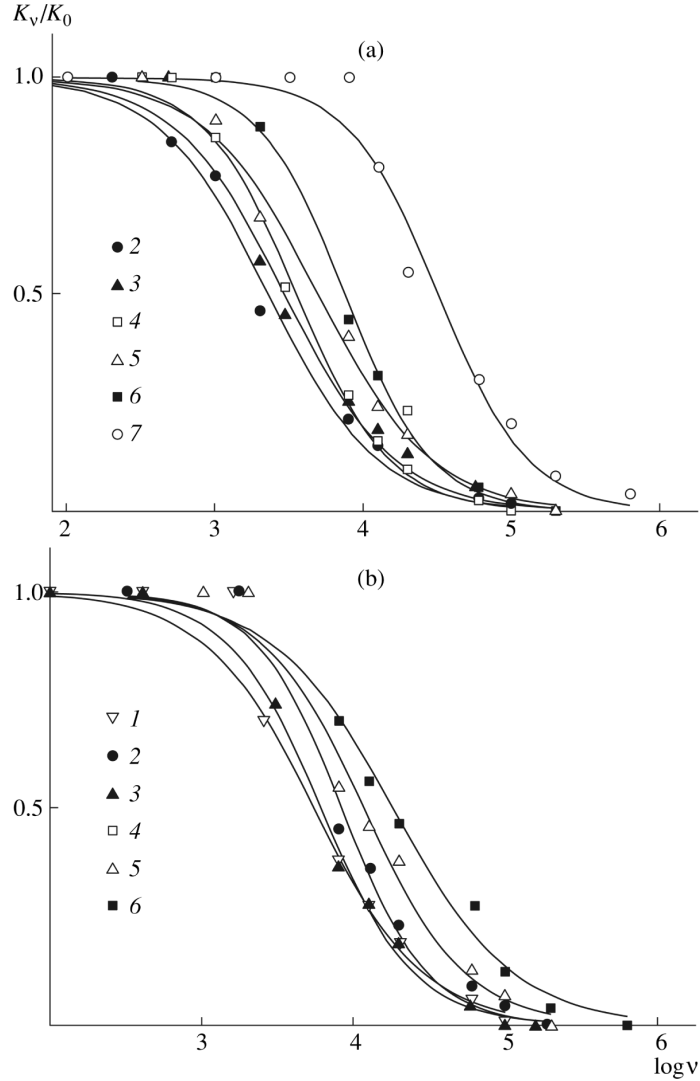}
\caption{Dispersion curves of specific Kerr constant for the polymers P3-1 (a) and P3-2 (b). The numbers of the curves correspond to the numeration of samples in Table 2.}
\label{09}
\end{figure}

Fig.~\ref{09} shows the electric birefringence curves measured for P3-1 and P3-2 polymers. As is seen, the dispersion dependences decline virtually to zero with growing electric field frequency and shift to the low-frequency range with growing molecular mass of the polymers. Such features of the dispersion dependences are characteristic of macromolecules that align to electric fields mostly according to the large-scale rotation mechanism.

On the basis of the dispersion dependences, relaxation times $\tau$ were estimated (Table 2). For the quantitative analysis, the relaxation times are compared to the molecular masses $M$ of polymers, their intrinsic viscosities $[\eta]$, and the viscosity of the solvent $\eta_0$ via relationship \cite{10,13}:
\begin{equation}
M[\eta]\eta_0D_r = FRT,
\end{equation}
where $D_r = 1/2 \tau$ is rotational diffusion coefficient of macromolecules. This comparison permits calculation of the model coefficient $F$ that characterizes the conformations of macromolecules. The values of $F$ are listed in Table 2. For the majority of samples, these values are within the limits predicted by the theory for kinetically rigid particles oriented in electric field according to the large-scale mechanism.

It is worth noting that the studied dendritic macromolecules contain polar groups only in the side branched substituents. In such a case, a question arises as to whether a macromolecule possesses the permanent (in the absence of the electric field) dipole moment or whether it appears only in the presence of the electric field due to a local reorientation of polar groups in the macromolecule. To answer this question, not only the dispersion of electric birefringence but also the decay of birefringence upon electric field switched off were examined for the sample P3-1-1 in non-polar dioxane. This experiment made it possible to determine the time of free relaxation of macromolecules. This time $(\tau_0 = (32 \pm 10) \times 10^{-6} s)$ turned out to be 2.5 times smaller than the time of the dispersion relaxation $\tau$. This result agrees well with the theoretically predicted factor equal to 3 for rigid polar molecules. The fact that the experimental ratio of relaxation times is somewhat smaller than the theoretical value is probably associated with kinetic flexibility of macromolecules at such high degrees of polymerization, as is also indicated by a high value of the model coefficient $F$. At the same time, it should be emphasized that the above result clearly testifies that the branched macromolecules with dendrons based on $L$-aspargic acid are characterized by a permanent dipole moment.

\subsection{Equilibrium Electric Birefringence}

\begin{figure}[!ht]
\centering
\includegraphics{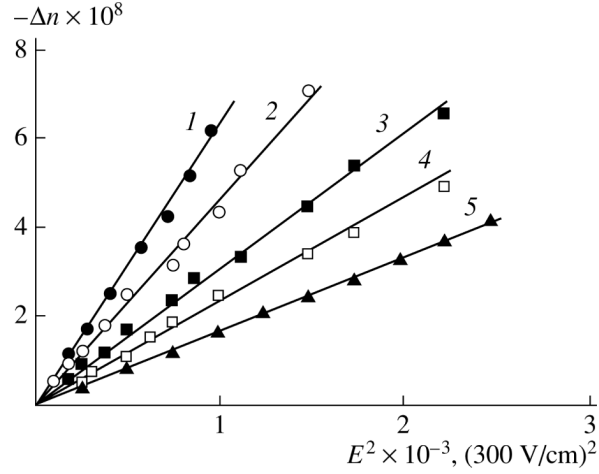}
\caption{Dependences of birefringence $\Delta n$ on the square of rectangular-pulsed electric field $E^2$ for the bromoform solution of polymer P3-2-5 at the concentrations $c \times 10^2 =$0.89~(1), 0.60~(2), 0.34~(3), and 0.17~(4) $g/cm^3$; (5) is the curve obtained for bromoform.}
\label{10}
\end{figure}

The results from equilibrium Kerr effect measurements for polymer P3-2-5 are shown in Fig.~\ref{10}, which plots electric birefringence $\Delta n$ as a function of the squared intensity $E^2$ of rectangular-pulsed electric field at various concentrations $c$ of the polymer solution.

The obtained dependences are straight lines. This facts indicates that the Kerr law is fulfilled and allows determination of the specific Kerr constants $K_c$ for the polymers in solution, from the relationship
\begin{equation}
K_c = \frac{\Delta n_{solution} - \Delta n_{solvent}}{cE^2},
\end{equation}
where $\Delta n_{solution}$ and $\Delta n_{solvent}$ are the electric birefringence of a solution and a solvent, respectively. For all the samples under study, the specific Kerr constants do not show a monotonic dependence on concentration. Therefore, the intrinsic Kerr constants $K = \lim_{c \to 0} K_c$ were obtained through averaging of $K_c$ values at various $c$. Table 2 shows the values of $K$ for the polymers under study.

\begin{figure}[!ht]
\centering
\includegraphics{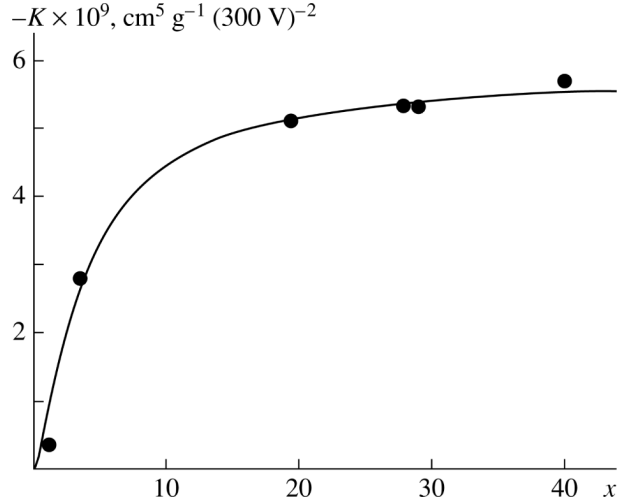}
\caption{The Kerr constant $K$ as a function of the reduced chain length $x$ for polymer P3-2. The closed circles denote experimental values, and the curve is the theoretical dependence calculated with the relationship (\ref{eq11}).}
\label{11}
\end{figure}

The Kerr constants of the polymers grow with increasing molecular masses. This tendency is typical of kinetically rigid macromolecules. Fig.~\ref{11} shows the experimental plot of the Kerr constant $K$ for the polymer P3-2 as a function of reduced chain length $x$,
\begin{equation}
x = \frac{M}{M_0}\frac{\lambda}{a},
\label{eqx}
\end{equation}
where $M$ is molecular mass of a sample, $M_0$ is molecular mass of the monomer unit, $\lambda$ is size of monomer unit projected on backbone, and $a$ is persistent length of the polymer.

The obtained set of experimental data, with polymerization degrees ranging in the interval 36\ldots{}1229, is compared with the theoretical curve for kinetically rigid wormlike molecules given by the relationship \cite{10,13}:

\begin{equation}
\frac{K}{K_{\infty}} = \frac{3 f^2_1(x)}{5f_2(x)} [1 - 0.6 tan^2 \theta \frac{f_1(x)}{f_2(x)}] \frac{\langle h^4 \rangle}{\langle h^2 \rangle}.
\label{eq11}
\end{equation}

Here, $\theta$ is angle between the dipole moment of the monomer unit $\mu_0$ and the direction of model chain growth, $\langle h^4 \rangle$ and $\langle h^2 \rangle$ are the fourth and the second moments of the length distribution function of vectors $h$ connecting the ends of persistent chains, and $K_{\infty}$ is the limiting value of the Kerr constant in the Gaussian region. The value of $K_{\infty}$ may be expressed as
\begin{equation}
K_{\infty} = (PQ)^2 2\pi N_A (n^2+2)^2S^2 \Delta a \frac{\mu_{0 \pa}^2}{135n(kT)^2M_0},
\end{equation}
where $P$ and $Q$ are multiplier factors of the internal field according to Onsager, $n$ is refractive index of solution, $S$ is number of monomer units in the Kuhn segment, and $\Delta a$ is the optical anisotropy of the monomer unit of mass $M_0$.

The theoretical curve provides the best description of the available experimental data in the case of the parameters $K_{\infty} = -59 \times 10^{-10} g^{-1} cm^5 (300 V)^{-2}$, $\theta = 0$, and $\lambda / a = 0.033$.

As noted above, the generalized wormlike chain model is preferable for the description of conformational and hydrodynamic properties of macromolecules of the P3-2 polymer in solutions \cite{14}. The persistent length of the polymer $a = 47~\AA$ is obtained if the size of monomer unit projected on backbone is taken $\lambda = 1.56~\AA$, as reported in \cite{9}. The value is in good agreement with $a = 53~\AA$ derived from molecular hydrodynamics experiments. This fact provides additional grounds in favor of the generalized wormlike chain model for description of the molecular characteristics of P3-2.

Since electric and dynamic birefringence measurements of the highest molecular mass sample of P3-1 were carried out in a non-polar solvent (dioxane), the longitudinal component of the dipole moment of the monomer unit $\mu_{0 \pa}$ in the chain growth direction can be calculated. This parameter is found to be $2.3~D$, which is considerably higher than $\mu_{0 \pa}$ reported for P1-1 $(0.8~D)$ \cite{7}. Thus, the introduction of rigid benzamide fragments into side dendrite substituents brings a significant increase in both the equilibrium rigidity and the dipole moment of dendritic macromolecules.

\section{\label{sec:conc} Conclusions}

Experimental results obtained in this work extend previous findings and give grounds for generalization that the conformational, hydrodynamic, optical, and dipole characteristics of hyperbranched macromolecules are determined not only by the generation number, but also by the structure of side dendrons.

The introduction of rigid benzamide fragments into side dendritic substituents significantly increases the equilibrium rigidity, the anisotropy of optical polarizability, and the dipole moment of macromolecules. The backbone conformation for the maximally elongated molecules of P3-2 is not planar and may be best described in terms of the generalized wormlike chain model.

The dendritic macromolecules with dendrons based on $L$-aspargic acid possess permanent dipole moment. In solvents that do not disturb intermolecular hydrogen bonds, the polymers exhibit a considerable kinetic rigidity in external fields.

\section{\label{sec:ack} Acknowledgements}
The work was supported by the Russian Foundation for Basic Research (projects 06-03-32601 and 06-03-32296) and by the Russian Federal Agency for Science and Innovations (project 2006-RI-190/001/225.)

\selectfont 

\end{document}